\documentclass[onecolumn,preprintnumbers,eqsecnum,superscriptaddress,12pt]{revtex4}
\usepackage{amsfonts}
\usepackage{amssymb}
\usepackage{amsmath}
\usepackage{epsf}
\usepackage{graphicx}
\usepackage{dcolumn}
\usepackage{bm}

\begin{document}

\title{Generalized Misner-Sharp Energy in $f(R)$ Gravity}

\author{Rong-Gen Cai}\email{cairg@itp.ac.cn}
\address{
Key Laboratory of Frontiers in Theoretical Physics, Institute of
Theoretical Physics, Chinese Academy of Sciences, P.O. Box 2735,
Beijing 100190, China}
\address{
Department of Physics, Kinki University, Higashi-Osaka, Osaka
577-8502, Japan}

\author{Li-Ming Cao}\email{caolm@itp.ac.cn}
\address{
Department of Physics, Kinki University, Higashi-Osaka, Osaka
577-8502, Japan}

\author{Ya-Peng Hu}\email{yapenghu@itp.ac.cn}
\address{
Key Laboratory of Frontiers in Theoretical Physics, Institute of
Theoretical Physics, Chinese Academy of Sciences, P.O. Box 2735,
Beijing 100190, China}
\address{
Graduate School of the Chinese Academy of Sciences, Beijing
100039, China}

\author{Nobuyoshi Ohta}\email{ohtan@phys.kindai.ac.jp}
\address{
Department of Physics, Kinki University, Higashi-Osaka, Osaka
577-8502, Japan}

\vspace*{1.cm}
\begin{abstract}
We study generalized Misner-Sharp energy in $f(R)$ gravity in a
spherically symmetric spacetime. We find that unlike the cases of
Einstein gravity and Gauss-Bonnet gravity, the existence of the
generalized Misner-Sharp energy depends on a constraint condition
in the $f(R)$ gravity. When the constraint condition is
satisfied, one can define a generalized Misner-Sharp energy, but
it cannot always be written in an explicit quasi-local form.
However, such a form can be obtained in a FRW universe and for
static spherically symmetric solutions with constant
scalar curvature. In the FRW universe, the generalized
Misner-Sharp energy is nothing but the total matter energy inside
a sphere with radius $r$, which acts as the boundary of a finite
region under consideration. The case of scalar-tensor gravity is
also briefly discussed.

\end{abstract}

\maketitle
\preprint{KU-TP 036}

\newpage

\section{Introduction}

A gravitational field has certainly an associated energy. However,
it is a rather difficult task to define energy for a gravitational
field in general relativity.  A local energy density of
gravitational field does not make any sense because the
energy-momentum pseudo-tensor of gravitational field, which
explicitly depends on metric and its first derivative, will vanish
due to the strong equivalence principle at any point of spacetime
in a locally flat coordinate~\cite{NonLocal1,NonLocal2,Szabados}.
In general relativity, however, there exist two well-known
definitions of total energy; one is  the Arnowitt-Deser-Misner
(ADM) energy $E_{ADM}$ at spatial infinity~\cite{ADM}, and the
other is the Bondi-Sachs (BS) energy $E_{BS}$ at null
infinity~\cite{Bondi} describing an isolated system in an
asymptotically flat spacetime.

Due to the absence of the local energy density of gravitational
field, it is tempting to define some meaningful quasi-local
energy, which is defined on a boundary of a given region in
spacetime. Indeed, it is possible to properly define such
quasi-local energies. Some useful definitions for quasi-local
energy exist in the literature, for instance, Brown-York
energy~\cite{York}, Misner-Sharp energy~\cite{Misner},
Hawking-Hayward energy~\cite{hawking,hayward} and Chen-Nester
energy~\cite{Chen}, etc. A nice review on this issue can be found
in~\cite{Szabados}. In this article, we focus on the Misner-Sharp
energy.

The Misner-Sharp energy $E$ is defined in a spherically symmetric
spacetime. Various properties of the Misner-Sharp energy are
discussed in some detail by Hayward in~\cite{Hayward,Hayward1}.
For example, the following properties are established. In the
Newtonian limit of a perfect fluid, the Misner-Sharp energy $E$
yields the Newtonian mass to leading order and the Newtonian
kinetic and potential energy in the next order. For test
particles, the corresponding Hajicek energy is conserved and has
the behavior appropriate to energy in the Newtonian and
special-relativistic limits. In the small-sphere limit, the
leading term in $E$ is the product of volume and the energy
density of the matter. In vacuo, the Misner-Sharp energy $E$
reduces to the Schwarzschild energy. At null and spatial infinity,
$E$ reduces to the BS and ADM energies, respectively. In
particular, it is shown that the conserved Kodama current produces
the conserved charge $E$.

In a four-dimensional, spherically-symmetric spacetime with metric
\begin{equation}
ds^2 = h_{ab} dx^adx^b + r^2 (x) d\Omega_2^2,
\end{equation}
where $a=0$, $1$, $x^a$ is the coordinate on a two-dimensional
spacetime $(M^2, h_{ab})$ and $d\Omega_2^2$ denotes the line
element for a two-dimensional sphere with unit radius, the
Misner-Sharp energy $E$ can be defined as
\begin{equation}
\label{1eq2} E(r) = \frac{r}{2G} \left(1-h^{ab}\partial_a
r\partial_b r\right).
\end{equation}
With this energy, the Einstein equations can be rewritten as
\begin{equation}
\label{1eq3} dE = A \Psi_a dx^a + W dV,
\end{equation}
where $A = 4\pi r^2$ is the area of the sphere with radius $r$ and
$V=4\pi r^3/3$ is its volume, $W$ is called  work density defined
as $W= -h^{ab}T_{ab}/2$ and $\Psi$ energy supply vector,
$\Psi_a=T_a^{\ b} \partial _b r +W \partial r_a$, with $T_{ab}$
being the projection of the four-dimensional energy-momentum tenor
$T_{\mu\nu}$ of matter in the normal direction of the
2-dimensional sphere.  The form (\ref{1eq3}) is called ``unified
first law"~\cite{Hayward2,Hayward3}. Projecting this form along a
trapping horizon, one is able to arrive at the first law of
thermodynamics for dynamical black hole
\begin{equation}
\langle dE,\xi \rangle =\frac{\kappa}{8\pi G}\langle dA,\xi
\rangle +W \langle dV,\xi \rangle,
\end{equation}
where $\xi$ is a projecting vector and $\kappa =
\frac{1}{2\sqrt{-h}}\partial_a (\sqrt{-h}h^{ab}\partial _br)$  is
surface gravity on the trapping horizon.  Defining $\delta Q =
\langle A\Psi,\xi \rangle  = T dS$, we can derive entropy formula
associated with apparent horizon in various gravity
theories~\cite{Cai-Cao1,Cai-Cao2,CCHK}. Indeed, the Minser-Sharp
energy plays an important role in connection between the Einstein
equations and first law of thermodynamics in FRW cosmological
setup~\cite{Cai-Cao1,Cai-Cao2,Cai-Kim,CCHK,AC07} and black hole
setup~\cite{Pad}.

Note that the original form (\ref{1eq2}) for the Misner-Sharp
energy is applicable  for  Einstein gravity without cosmological
constant in four dimensions, thus it is tempting to give
corresponding forms for the case with a cosmological constant
and/or in other gravity theories.  Indeed, a generalized form is
given for Gauss-Bonnet gravity and more general Lovelock gravity
in \cite{Maeda,Maeda1}. In particular, we would like to mention
here that Gong and Wang in \cite{GW} introduce a modified
Misner-Sharp energy and discuss its relation to horizon
thermodynamics. With the generalized Misner-Sharp energy, it is
shown that the Clausius relation $\delta Q=TdS$ indeed gives
correct entropy formula for Lovelock gravity~\cite{Cai-Cao1,CCHK}.

Recently, a kind of modified gravity theories, $f(R)$, whose
Lagrangian is a function of curvature scalar $R$, has attracted a
lot of attention. A main motivation is to explain the observed
accelerated expansion of the universe without introducing the
exotic dark energy with a large negative pressure. For a review on
$f(R)$ gravity, see \cite{Sotiriou}. Of course, $f(R)$ gravity is
a simple generalization of Einstein gravity; when $f(R)=R$, it
goes back to Einstein theory.  However, $f(R)$ is quite different
from another generalization of Einstein gravity, Lovelock gravity.
The equations of motion of the latter do not contain more than
second-order derivatives, while the equations of motion for the
former do. In addition, let us notice that in some sense, the
$f(R)$ gravity is quite similar to scalar-tensor gravity, a
generalization of Einstein gravity again.

In this paper we are mainly concerned with the question whether
there exists a similar Misner-Sharp energy for $f(R)$ gravity in a
spherically symmetric spacetime.  For this goal, we will take two
methods, which are basically equivalent, in fact. One is called
integration method, and the other is conserved charge method
associated with the Kodama current.  The integration method is
introduced in a previous paper of ours~\cite{CCHK} for the case of
radiation matter in Lovelock gravity. We find that existence of a
generalized Misner-Sharp energy is not trivial for $f(R)$ gravity.
Its existence depends on a constraint. Once the constraint is
satisfied, we could have a generalized Misner-Sharp energy.
Otherwise, the answer is negative. The same situation happens for
the scalar-tensor gravity theory.

The organization of the  paper is as follows. In Sec.~II, as a
warm-up exercise, we derive the generalized Misner-Sharp energy in
Gauss-Bonnet gravity by using the integration method and by
generalizing the discussion in \cite{CCHK} to more general matter
content.  In Sec.~III, we discuss the generalized Misner-Sharp
energy in $f(R)$ gravity by the integration method and conserved
charge method, respectively. Sec.~IV is devoted to investigating
some special cases, homogeneous and isotropic FRW cosmology and
static spherically symmetric case. In these cases the generalized
Misner-Sharp energy has a simple form.  The conclusion and some
discussions are given in Sec.~V. In the appendix, we briefly
discuss the generalized Misner-Sharp energy for scalar-tensor
gravity in a FRW universe.

\section{Generalized Misner-Sharp energy in Gauss-Bonnet gravity: integration method}

The equations of motion  of Gauss-Bonnet gravity can be written
down as
\begin{equation}
G_{\mu \nu }+\alpha H_{\mu \nu }+\Lambda g_{\mu \nu }=8\pi G
T_{\mu \nu}, \label{Equation of GBG}
\end{equation}%
where
\begin{eqnarray}
G_{\mu \nu }&=&R_{\mu \nu }-\frac{1}{2}Rg_{\mu \nu }, \nonumber\\
H_{\mu \nu}&=&2(RR_{\mu \nu }-2R_{\mu \alpha }R_{\nu }^{\ \alpha
}-2R^{\alpha \beta }R_{\mu \alpha \nu \beta } +R_{\mu }^{\ \alpha
\beta\gamma }R_{\nu \alpha \beta \gamma
})-\frac{1}{2}g_{\mu\nu}L_{GB},
\end{eqnarray}
and $\alpha $\ is a coupling constant with dimension of length squared.
The Gauss-Bonnet term is $L_{GB}=R^{2}-4R_{\mu \nu }R^{\mu \nu
}+R_{\mu \nu \rho \sigma }R^{\mu \nu \rho \sigma }$.

Consider an $n$-dimensional spherically symmetric spacetime of
metric in the double-null form
\begin{equation}
ds^{2}=-2e^{-\varphi(u,v)}dudv+r^{2}(u,v)\gamma _{ij}dz^{i}dz^{j},
\label{2eq3}
\end{equation}%
where $\gamma _{ij}$ is the metric on an $(n-2)$-dimensional
constant curvature space $K^{n-2}$ with its sectional curvature
$k=\pm 1,0$, and the two-dimensional spacetime spanned by two null
coordinates $(u,v)$ and its metric are denoted as $(M^{2}, h_{ab})$. Thus, the
equations of gravitational field (\ref{Equation of GBG}) can be
written explicitly as~\cite{Maeda1}
\begin{eqnarray}
-\frac{8\pi
G}{n-2}rT_{uu}&=&(r_{,uu}+\varphi_{,u}r_{,u})\Big[1+\frac{2{\tilde
{\alpha }}}{r^{2}}(k+2e^{\varphi}r,_{u}r,_{v})\Big] , \nonumber\\
-\frac{8\pi G}{n-2}rT_{vv}&=&(r,_{vv}+\varphi,_{v}r,_{v})\Big[
 1+\frac{2{\tilde {\alpha }}}{r^{2}}%
(k+2e^{\varphi}r,_{u}r,_{v})\Big], \notag \\
\frac{8\pi
G}{n-2}r^{2}T_{uv}&=&rr,_{uv}+(n-3)r,_{u}r,_{v}+\frac{n-3}{2}ke^{-\varphi}
+\frac{{\tilde {\alpha }}} {2r^{2}} [(n-5)k^{2}e^{-\varphi}
+4rr,_{uv}(k+2e^{\varphi}r,_{u}r,_{v})\nonumber\\
&&+4(n-5)r,_{u}r,_{v}(k+e^{\varphi}r,_{u}r,_{v})] -
\frac{n-1}{2}{\tilde {\Lambda }}r^{2}e^{-\varphi},
 \label{2eq4}
\end{eqnarray}%
where ${\tilde {\alpha}}=(n-3)(n-4)\alpha ,$ ${\tilde {\Lambda
}}=2\Lambda /[(n-1)(n-2)].$

The essential point of the integration method is that, similar to
the case of Einstein gravity (\ref{1eq3}), one assumes the
equations (\ref{2eq4}) of gravitational field can be cast into the form
\begin{equation}
dE_{eff}=A\Psi_a dx^a +WdV, \label{2eq5}
\end{equation}
where $A=V_{n-2}^{k}r^{n-2}$ and $V =V_{n-2}^{k}r^{n-1}/(n-1)$ are
area and volume of the $(n-2)$-dimensional space with radius $r$,
and energy supply  vector $\Psi $ and energy density $W$ are
defined on $(M^{2},h_{ab})$ as in the case of Einstein gravity.
The right hand side in (\ref{2eq5}) can be explicitly expressed as
\begin{equation}
A\Psi_a dx^a +WdV=A(u,v)du+B(u,v)dv,
 \label{2eq6}
\end{equation}%
where
\begin{eqnarray}
A(u,v) &=&V_{n-2}^{k}r^{n-2}e^{\varphi}(r,_{u}T_{uv}-r,_{v}T_{uu}), \\
B(u,v)
&=&V_{n-2}^{k}r^{n-2}e^{\varphi}(r,_{v}T_{uv}-r,_{u}T_{vv}).
\label{Components}
\end{eqnarray}
With the equations in (\ref{2eq4}), we can express $A$ and $B$ in
terms of geometric quantities as
\begin{eqnarray}
A(u,v) &=&\frac{V_{n-2}^{k}}{8\pi
G}e^{\varphi}(n-2)r^{n-4}\Big\{\frac{e^{-\varphi}
}{2r^{2}}r,_{u}[-(n-1){\tilde {\Lambda
}}%
r^{4}+(n-3)r^{2}(k+2e^{\varphi}r,_{u}r,_{v}) \notag\\
&&+(n-5){\tilde{\alpha }}%
(k+2e^{\varphi}r,_{u}r,_{v})^{2}+2e^{\varphi}r^{3}r,_{uv}+4e^{\varphi}{\tilde
{\alpha }}r(k+2e^{\varphi}r,_{u}r,_{v})r,_{uv}]\notag \\
&&+rr,_{v}\Big[1+\frac{2{\tilde {\alpha }}}{r^{2}}(k+2e^{\varphi}r,_{u}r,_{v})\Big]
(\varphi,_{u}r,_{u}+r,_{uu})\Big\},
\notag \\
B(u,v) &=&\frac{V_{n-2}^{k}}{8\pi G}e^{\varphi}(n-2)r^{n-4}\Big\{\frac{e^{-\varphi}%
}{2r^{2}}r,_{v}[-(n-1){\tilde {\Lambda
}}%
r^{4}+(n-3)r^{2}(k+2e^{\varphi}r,_{u}r,_{v}) \notag\\
&&+(n-5){\tilde {\alpha }}
(k+2e^{\varphi}r,_{u}r,_{v})^{2}+2e^{\varphi}r^{3}r,_{uv}+4{\tilde
{\alpha }}re^{\varphi}(k+2e^{\varphi}r,_{u}r,_{v})r,_{uv}] \notag \\
&&+rr,_{u}\Big[1+\frac{2{\tilde {\alpha
}}}{r^{2}}(k+2e^{\varphi}r,_{u}r,_{v})\Big](\varphi,_{v}r,_{v}+r,_{vv})\Big\}.
\label{2eq9}
\end{eqnarray}%
Now we try to derive the generalized Misner-Sharp energy by
integrating the equation (\ref{2eq5}).  Clearly, if it is
integrable, the following integrable condition has to be satisfied
\begin{equation}
\frac{\partial A(u,v)}{\partial v}=\frac{\partial B(u,v)}{\partial
u}.
 \label{2eq10}
\end{equation}%
It is easy to check that $A$ and $B$ given in (\ref{2eq9}) indeed
satisfy the integrable condition (\ref{2eq10}). Thus directly
integrating (\ref{2eq5}) gives the generalizing Misner-Sharp
energy
\begin{eqnarray}
E_{eff} &=&\int A(u,v)du +\int\Big[B(u,v)-\frac{\partial}{\partial v}\int A(u,v)du
 \Big]dv \notag \\
&=&\frac{(n-2)V_{n-2}^{k}r^{n-3}}{16\pi G}[-%
{\tilde {\Lambda }}r^{2}+(k+2e^{\varphi}r,_{u}r,_{v})+{\tilde
{\alpha }}r^{-2}(k+2e^{\varphi}r,_{u}r,_{v})^{2}]. \label{2eq11}
\end{eqnarray}%
Note that here the second term in the first line of (\ref{2eq11})
in fact vanishes and we have fixed an integration constant so that
$E_{eff}$ reduces to the Misner-Sharp energy in Einstein gravity
when ${\tilde{\alpha }}=0$.
In addition, the generalized Misner-Sharp energy
can  be rewritten in a covariant form
\begin{eqnarray}
\label{2eq12}
E_{eff}&=&\frac{(n-2)V_{n-2}^{k}r^{n-3}}{16\pi G }[-%
{\tilde {\Lambda }}r^{2}+(k-h^{ab}D_{a}rD_{b}r)+{\tilde {\alpha
}}r^{-2}(k-h^{ab}D_{a}rD_{b}r)^{2}].
\end{eqnarray}%
This is the generalized Misner-Sharp energy given by Maeda and Nozawa in
\cite{Maeda1} through Kodama conserved charge method.

\section{Generalized Misner-Sharp energy in $f(R)$ gravity: the general case}

In this section, we first try to derive the generalized
Misner-Sharp energy in $f(R)$ gravity by using the integration
method. Then we consider the conserved charge method.  Here we
consider the four-dimensional case with spherical symmetry, and
the line element is
\begin{equation}
ds^{2}=-2 e^{-\varphi(u,v)}dudv+r^{2}(u,v)(d\theta ^{2}+\sin
^{2}\theta d\phi ^{2}).  \label{3eq1}
\end{equation}%
The action of the $f(R)$ gravity in the metric formalism is
\begin{equation}
S=\frac{1}{16\pi G }\int d^{4}x\sqrt{-g}f(R)+S_{matter},
\label{3eq2}
\end{equation}%
Varying the action with respect to metric yields equations of
gravitational field
\begin{equation}
f_{R}R_{\mu \nu }-\frac{1}{2}fg_{\mu \nu }-\nabla _{\mu }\nabla
_{\nu }f_{R}+g_{\mu \nu }\square f_{R}=8\pi G T_{\mu \nu },
\label{3eq3}
\end{equation}%
where $f_{R}=df(R)/dR,$ and $T_{\mu \nu }$\ is the energy-momentum
tensor for matter field  from $S_{matter}$. Note that the field
equations also can be rewritten in the form
\begin{equation}
G_{\mu \nu }\equiv R_{\mu \nu }-\frac{1}{2}g_{\mu \nu }R=\frac{1}{f_{R}}\Big[%
\frac{1}{2}g_{\mu \nu }(f-Rf_{R})+\nabla _{\mu }\nabla _{\nu
}f_{R}-g_{\mu \nu }\square f_{R}\Big]
 +\frac{8\pi G}{f_{R} }T_{\mu \nu}. \label{3eq4}
\end{equation}%
In this case, the right hand side can be regarded as an effective
energy-momentum tensor.

\subsection{Integration method}

In the metric (\ref{3eq1}), the  field equations (\ref{3eq3}) can
be explicitly expressed as
\begin{eqnarray}
8\pi GT_{uu}&=& -2f_{R}\frac{\varphi,_{u}r,_{u}+r,_{uu}}{r}-f_{R},_{uu}
-\varphi,_{u}f_{R},_{u},  \notag \\
8\pi G T_{vv}&=& -2f_{R}\frac{\varphi,_{v}r,_{v}+r,_{vv}}{r}-f_{R},_{vv}
-\varphi,_{v}f_{R},_{v},  \notag \\
8\pi G T_{uv}&=& f_{R}\varphi_{,uv}-2f_{R}\frac{r,_{uv}}{r}+\frac{1}{2%
}fe^{-\varphi}+f_{R},_{uv}+\frac{2r,_{u}f_{R},_{v}+2r,_{v}f_{R},_{u}}{r}.
\label{3eq5}
\end{eqnarray}%
In this case, following the method discussed in the previous
section, we  obtain
\begin{eqnarray}
A(u,v) &=& 4\pi r^{2} e^{\varphi}(r,_{u}T_{uv}-r,_{v}T_{uu})  \notag \\
&=&\frac{r^{2}e^{\varphi}}{2G}\Big[r,_{u}\Big(f_{R}\varphi_{,uv}
-2f_{R}\frac{r,_{uv}}{r}+\frac{1}{2}fe^{-\varphi}+f_{R},_{uv}+\frac{%
2r,_{u}f_{R},_{v}+2r,_{v}f_{R},_{u}}{r}\Big) \notag \\
&&+ r,_{v}\Big(2f_{R}\frac{\varphi,_{u}r,_{u}+r,_{uu}}{r}+f_{R},_{uu}+
\varphi,_{u}f_{R},_{u}\Big)\Big],  \notag \\
B(u,v) &=&4\pi r^{2}e^{\varphi}(r,_{v}T_{uv}-r,_{u}T_{vv})  \notag \\
&=&\frac{r^{2}e^{\varphi}}{2G}\Big[r,_{v}\Big(f_{R}\varphi_{,uv}
-f_{R}\frac{2r,_{uv}}{r}+\frac{1}{2}fe^{-\varphi}+f_{R},_{uv}+\frac{%
2r,_{u}f_{R},_{v}+2r,_{v}f_{R},_{u}}{r}\Big) \notag \\
&&+r,_{u}\Big(2f_{R}\frac{\varphi,_{v}r,_{v}+r,_{vv}}{r}+f_{R},_{vv}+
\varphi,_{v}f_{R},_{v}\Big)\Big]. \label{3eq6}
\end{eqnarray}
Checking the integrable condition, however, unlike the case of
Gauss-Bonnet gravity, we find that  it is not always satisfied for
the $f(R)$ gravity:
\begin{eqnarray}
\frac{\partial A(u,v)}{\partial v}-\frac{\partial B(u,v)}{\partial
u}&=&- r^{2}e^{\varphi} [(\varphi,_{u}r,_{u}+
r,_{uu})(f_{R},_{vv}+\varphi,_{v}f_{R},_{v})
-(\varphi,_{v}r,_{v}+r,_{vv})  \notag \\
&&(f_{R},_{uu}+\varphi,_{u}f_{R},_{u})]/2G \label{3eq7}.
\end{eqnarray}%
If the right hand side of the above equation vanishes, in
principle, one is able to obtain a generalized Misner-Sharp energy
by integrating (\ref{2eq5}). On the other hand, if the integrable
condition is not satisfied, one is not able to rewrite the form
$Adu+Bdv$ as a total differential form, which implies that
generalized Misner-Sharp energy does not exist in this case.  Now
we assume that the integrable condition is satisfied, that is to
say, the right hand side of the equation (\ref{3eq7}) vanishes.
Thus, we can obtain the generalized Misner-Sharp energy for the
$f(R)$ gravity as
\begin{eqnarray}
E_{eff}\ &=&\int A(u,v)du +\int\Big[B(u,v)-\frac{\partial}{\partial v}
\int A(u,v)du \Big]dv \nonumber \\
&=&\frac{r}{2G}\Big[(1+2 e^{\varphi} r,_{u}r,_{v})f_{R}+%
\frac{1}{6}r^{2}(f-f_{R}R) + r e^{\varphi}%
(f_{R},_{u}r,_{v}+f_{R},_{v}r,_{u})\Big]  \notag \\
&&-\frac{1}{2G}\int \Big[f_{R},_{u}e^{\varphi}(r^{2}r,_{v}),_{u}+f_{R},_{u}(r-%
\frac{1}{6}r^{3}R)+f_{R},_{v}r^{2}(r,_{u} e^{\varphi}),_{u}\Big]du  \notag \\
&=&\frac{r}{2G}\Big[(1-h^{ab}\partial _{a}r\partial_{b}r)f_{R}+\frac{1}{6}%
r^{2}(f-f_{R}R)- rh^{ab}\partial _{a}f_{R}\partial _{b}r\Big] \notag
\\
&&-\frac{1}{2G}\int \Big[f_{R},_{u}e^{\varphi}(r^{2}r,_{v}),_{u}+f_{R},_{u}\Big(r-%
\frac{1}{6}r^{3}R\Big)+f_{R},_{v}r^{2}(r,_{u} e^{\varphi}),_{u}\Big]du,
\label{3eq8}
\end{eqnarray}%
where we have used
$R=2[\frac{1}{r}+e^\varphi(2\frac{r_{,u}r_{,v}}{r^2}-\varphi_{,uv}+4\frac{r_{,uv}}{r})]$
and $f_{,u}=f_R R_{,u}$. We see that $E_{eff}$ reduces to the
Misner-Sharp energy in the Einstein gravity when $f_R=1$.
Unfortunately, we see from (\ref{3eq8}) that due to the existence
of the integration in (\ref{3eq8}), we cannot arrive at an
explicit quasi-local energy for the general case.  In the next
section, however, we will show that the integration can be carried
out in some special cases. Before that, we will first obtain the
same result using the conserved charge method in the next
subsection.

\subsection{Conserved charge method}

In a spherically symmetric spacetime, one can define a Kodama
vector. The energy-momentum tensor together with the Kodama vector
can lead to a conserved current, whose corresponding conserved
charge is just the Misner-Sharp energy in Einstein gravity. In
Gauss-Bonnet gravity, Maeda and Nozawa~\cite{Maeda1} obtain the
generalized Misner-Sharp energy with help of this method. In this
section we would like to see whether the conserved current method
leads to a generalized Misner-Sharp energy for the $f(R)$ gravity.

For a spherically symmetric spacetime, one can define the Kodama
vector as~\cite{Kodama, Sasaki}
\begin{equation}
K^{\mu }=-\epsilon ^{\mu \nu }\nabla _{\nu}r,
\label{3eq9}
\end{equation}%
where $\epsilon _{\mu \nu }=\epsilon _{ab}(dx^{a})_{\mu }(dx^{b})_{\nu}$, and $%
\epsilon _{ab}$ is the volume element of $(M^{2},h_{ab})$. For the
spherically symmetric spacetime  (\ref{3eq1}), we have
\begin{equation}
K^{\mu}=e^{\varphi}r,_{v}\Big(\frac{\partial }{\partial
u}\Big)^{\mu} -e^{\varphi}r,_{u}\Big(\frac{\partial }{\partial
v}\Big)^{\mu}. \label{3eq10}
\end{equation}
Conservation of the energy-momentum tensor for matter fields
$T_{\mu \nu }$ \ in (\ref{3eq3}) guarantees that the left hand
side of the equation (\ref{3eq3}) is also divergence-free, which
can be easily checked by using the identity
\begin{equation}
(\square \nabla _{\nu}-\nabla _{\nu}\square )F=R_{\mu \nu }\nabla
^{\mu }F, \label{3eq11}
\end{equation}%
where $F$ is an arbitrary scalar function. With the Kodama vector,
define an  energy current as
\begin{equation}
J^{\mu }=-T_{\nu}^{\mu }K^{\nu}.  \label{3eq12}
\end{equation}%
However, we find that unlike in the cases of Einstein gravity and
Gauss-Bonnet gravity~\cite{Maeda1}, the energy current defined in
(\ref{3eq12}) is  not always divergence-free for the $f(R)$
gravity except the case with condition
\begin{equation}
\nabla _{\mu }\nabla _{\nu}f_{R}\nabla ^{\mu }K^{\nu}=0.
\label{3eq13}
\end{equation}%
Namely, if the constraint equation (\ref{3eq13}) is satisfied, the
energy current is divergence-free
\begin{equation}
\nabla _{\mu }J^{\mu }=0.
\end{equation}%
In this case,  we can define an associated conserved charge
\begin{equation}
Q_{J}=\int_{\Sigma }J^{\mu }d\Sigma _{\mu }, \label{3eq15}
\end{equation}%
where $\Sigma $ is some hypersurface and $d\Sigma _{\mu }=\sqrt{-g}%
dx^{v}dx^{\lambda }dx^{\rho }\delta _{\mu v\lambda \rho }$ is a
directed surface line element on $\Sigma$. By using the line
element in (\ref{3eq1}) and equations in (\ref{3eq5}), we obtain
\begin{eqnarray}
Q_{J}&=&\int_{\Sigma }J^{\mu }d\Sigma _{\mu } \notag \\
&=&\frac{r}{2G}\Big[(1-h^{ab}\partial _{a}r\partial_{b}r)f_{R}
+\frac{1}{6}r^{2}(f-f_{R}R)-rh^{ab}\partial _{a}f_{R}\partial_{b}r\Big] \notag \\
&&-\frac{1}{2G}\int
\Big[f_{R},_{u}e^{\varphi}(r^{2}r,_{v}),_{u}+f_{R},_{u}\Big(r-
\frac{1}{6}r^{3}R\Big)+f_{R},_{v}r^{2}(r,_{u}e^{\varphi}),_{u}\Big]du ,
\label{3eq16}
\end{eqnarray}%
where we have chosen the hypersurface $\Sigma$ with a given $v$.
One can immediately see that the charge $Q_{J}$ is precisely
the generalized Misner-Sharp energy $E_{eff}$ given by the
integration method in~(\ref{3eq8}). Again, this is not a satisfying
situation since we cannot express the generalized Misner-Sharp
energy in a true quasi-local form.

\section{Generalized Misner-Sharp energy in f(R) gravity: special cases}

The existence of the integration in (\ref{3eq8}) is painful. An
interesting question is whether it will be absent in some special
cases. The answer is positive. We will here discuss two special
cases. One is the homogeneous and isotropic FRW universe and the
other is the static spherically symmetric spacetime with constant
scalar curvature.

\subsection{FRW Universe}

 Consider the metric
\begin{equation}
ds^{2}=-dt^{2}+e^{2\psi (t,\rho )}d\rho ^{2}+r^{2}(t,\rho
)(d\theta ^{2}+\sin ^{2}\theta d\phi ^{2}).  \label{4eq1}
\end{equation}%
In this metric, the Kodama vector  is
\begin{equation}
K^{a}=e^{-\psi }r,_{\rho }\Big(\frac{\partial }{\partial t}\Big)^{a}
-e^{-\psi }r,_{t}\Big(\frac{\partial }{\partial \rho }\Big)^{a}.
\label{4eq2}
\end{equation}%
Following the same procedure, we can rewrite the equation in
(\ref{2eq6}) as
\begin{equation}
A\Psi_a dx^a +WdV=A(t,\rho )dt+B(t,\rho )d\rho. \label{4eq3}
\end{equation}%
where
\begin{eqnarray}
A(t,\rho ) &=&4\pi r^{2}e^{-2\psi }(T_{t\rho }r,_{\rho }-T_{\rho \rho
}r,_{t}),  \notag \\
B(t,\rho ) &=&4\pi r^{2}(T_{tt}r,_{\rho }-T_{t\rho }r,_{t}).
\label{4eq4}
\end{eqnarray}%
With the equations of gravitational field of the $f(R)$ gravity,
$A$ and $B$ can be expressed in terms of geometric quantities.
One can then arrive at the generalized Misner-Sharp energy
\begin{eqnarray}
E_{eff} &=&\int B(t,\rho )d\rho +\int\Big[A(t,\rho )-\frac{\partial}{\partial t}
\int B(t,\rho )d\rho \Big]dt \notag\\
&=& \frac{r}{2G} \Big[(1-h^{ab}\partial
_{a}r\partial _{b}r)f_{R}+\frac{r^2}{6}(f-f_{R}R)-r %
h^{ab}\partial _{a}f_{R}\partial _{b}r \Big] \notag \\
&&+\frac{1}{2G}\int \Big\{f_{R,_\rho}\Big[(-e^{-2\psi
}r^{2}r,_{\rho }\psi ,_{\rho }+e^{-2\psi }r^{2}r,_{\rho \rho
}-r^{2}r,_{t}\psi ,_{t})-r(1+r,_{t}^{2}-e^{-2\psi }r,_{\rho
}^{2})+\frac{1}{6}r^{3}R\Big]
\notag\\
&& \qquad +r^{2}f_{R,t}(\psi ,_{t}r,_{\rho }-r,_{t\rho
})\Big\}d\rho, \label{4eq5}
\end{eqnarray}%
where  the integrable condition is assumed to be satisfied
\begin{equation}
\frac{\partial A(t,\rho )}{\partial \rho }-\frac{\partial B(t,\rho )}{%
\partial t}=0\text{ }.  \label{4eq6}
\end{equation}%
Now we express the generalized Misner-Sharp energy in a FRW metric
\begin{equation}
ds^{2}=-dt^{2}+\frac{a^{2}(t)d\rho ^{2}}{1-k\rho ^{2}}+r^2(t,\rho
)(d\theta ^{2}+\sin ^{2}\theta d\phi ^{2}), \label{FRWMetric}
\end{equation}%
where $r(t,\rho )\equiv a(t)\rho $, and $e^{\psi (t,\rho
)}=\frac{a(t)}{\sqrt{1-k\rho ^{2}}}$ corresponding
to~(\ref{4eq1}).
Because in the FRW universe, the Ricci scalar
$R=6(\frac{k}{a^{2}}+\frac{\overset{.}{a}^{2}}{a^{2}}+\frac{%
\overset{..}{a}}{a})$ just depends on time, we can check that the
integrand in the final step in~(\ref{4eq5}) exactly vanishes.
Thus, the generalized Misner-Sharp energy $E_{eff}$\ in this case
can be explicitly expressed as
\begin{eqnarray}
E_{eff} &=&\frac{r}{2G}\Big[(1-h^{ab}\partial _{a}r\partial _{b}r)f_{R}+\frac{1}{6}%
r^{2}(f-f_{R}R)-rh^{ab}\partial _{a}f_{R}\partial
_{b}r\Big]  \notag \\
&=&\frac{%
r^{3}}{2G} \left
(\frac{1}{r_{A}^{2}}f_{R}+\frac{1}{6}(f-f_{R}R)+H\partial
_{t}f_{R} \right),  \label{4eq8}
\end{eqnarray}%
where $r_{A}=1/\sqrt{H^{2}+\frac{k}{a^{2}}}$, which in fact, is
the location of apparent horizon of the FRW universe.

\subsection{Static spherically symmetric case}

The general line element of a static spherically symmetric
spacetime can be written down as
\begin{equation}
ds^{2}=-\lambda(r)dt^{2}+g(r)dr^{2}+r^{2}d\Omega _{2}^{2},
\end{equation}
where $\lambda $ and $g$ are two functions of the radial
coordinate. In this case, the Kodama vector is
\begin{equation}
K^{\mu} =\frac{1}{\sqrt{g\lambda}}\left (\frac{\partial }{\partial t}\right )^{\mu}. \\
\end{equation}
Using the static spherically symmetric metric, we can easily check
that the constraint (\ref{3eq13}) is naturally satisfied
\begin{equation}
\nabla _{\mu }\nabla _{\nu }f_{R}\nabla ^{\mu }K^{\nu} =0. \\
\end{equation}
Following the same procedure, we can rewrite the equation in
(\ref{2eq6}) as
\begin{equation}
\label{4eq12} dE_{eff}  =A(t,r)dt+B(t,r)dr,
\end{equation}
where
\begin{eqnarray}
A(t,r) &=&\frac{4\pi r^{2}}{g}(T_{tr }r,_{r}-T_{r r }r,_{t})=0, \\
B(t,r) &=&\frac{4\pi r^{2}}{\lambda}(T_{tt}r,_{r}-T_{t r }r,_{t}) \notag \\
&=&\frac{r^{2}}{2 G} \left
(\frac{1}{2}(f-f_{R}R)+\frac{1}{r^2}(1+\frac{rg^{'}}{g^{2}}-\frac{1}{g})f_{R}
+f_{R,r}(\frac{g^{'}}{2g^{2}}-\frac{2}{r
g})-\frac{1}{g}f_{R,rr}\right),
\end{eqnarray}
where a prime denotes the derivative with respect to $r$.
Integrating (\ref{4eq12}) gives the generalized Misner-Sharp
energy
\begin{eqnarray}
\label{4eq15}
 E_{eff} &=&\int
B(t,r)dr=\frac{r}{2G}\left((1-h^{ab}\partial _{a}r\partial
_{b}r)f_{R}+\frac{r^2}{6}(f-f_{R}R)- r h^{ab}\partial
_{a}f_{R}\partial _{b}r \right )  \notag \\
&&-\frac{1}{2G}\int dr
(\frac{r^{2}g^{'}}{2g^{2}}+r-\frac{r}{g}-\frac{1}{6 }r^{3}R)
f_{R,r}.
\end{eqnarray}
Clearly the integral in (\ref{4eq15}) will be absent in two cases,
one is $ \frac{r^{2}g^{'}}{2g^{2}}+r-\frac{r}{g}-\frac{1}{6
}r^{3}R=0$, the other is $f_{R,r}=0$.  We here consider the latter
case. The trivial case with $f(R)=R$ naturally satisfies the
condition. In this case, $f_R=1$, and (\ref{4eq15}) gives the
Misner-Sharp energy. A little nontrivial case is that the solution
is a constant curvature one with scalar curvature $R=R_0 =const.$
In that case, $f_{R,r}=0$, and (\ref{4eq15}) reduces to
\begin{equation}
 E_{eff} =\frac{r}{2G}\left((1-h^{ab}\partial _{a}r\partial
_{b}r)f_{R}+\frac{r^2}{6}(f-f_{R}R)\right ).
\end{equation}
Note that here  $R$, $f_R$ and $f$ are all constants. Compare to
the Misner-Sharp energy (\ref{1eq2}) and the (\ref{2eq12}),  we
can see clearly that this  expression is nothing but the
generalized Misner-Sharp energy with a cosmological constant. Here
the effective Newtonian constant is $G/f_R$ and the effective
cosmological constant $\Lambda = -(f-Rf_R)/(2f_R)$.

\section{Conclusion and discussion}

The Misner-Sharp quasi-local energy plays a key role in
understanding the ``unified first law", the relation between the
first law of thermodynamics and dynamical equations of
gravitational field and thermodynamics of apparent horizon in FRW
universe, etc. In this paper we studied the generalized
Misner-Sharp energy in $f(R)$ gravity by two approaches. One is
the integration method and the other is the conserved charge
method. It turns out that in general we cannot arrive at an
explicit expression for the generalized Misner-Sharp energy in a
quasi-local form [see (\ref{3eq8}) and (\ref{4eq5})], even
assuming the integrable condition (\ref{2eq10}) is satisfied. This
situation is quite different from the cases of Einstein gravity
and Gauss-Bonnet gravity.  This is certainly related to the fact
that for the $f(R)$ gravity, the energy current (\ref{3eq12}) is
not always divergence-free, while it does in Einstein and
Gauss-Bonnet gravities. The existence of the conserved current
requires (\ref{3eq13}) is satisfied.

Some remarks on our results are in order.

(1) The relation between the two methods to derive the generalized
Misner-Sharp energy.  We obtained the same  generalized
Misner-Sharp energy by employing two  methods: integration and
conserved charge methods.  At first glance, these two methods
looks different, but in fact, they are equivalent. First let us
notice that the constraint equation (\ref{3eq7}) has a relation to
the one (\ref{3eq13}):
\begin{equation}
\frac{\partial A(u,v)}{\partial v}-\frac{\partial B(u,v)}{\partial u}%
=-e^{-\varphi}r^{2}\nabla _{\mu }\nabla _{v}f_{R}\nabla ^{\mu
}K^{v}/2. \label{5eq1}
\end{equation}%
Namely these two integrable conditions are equivalent. Second,
 substituting the conserved current in (\ref{3eq12})
into (\ref{3eq15}), we can write the associated charge
\begin{eqnarray}
Q_{J}&=&\int_{\Sigma }J^{\mu }d\Sigma _{\mu } \notag \\
&=&\int 4\pi r^{2}e^{\varphi}(r,_{u}T_{uv}-r,_{v}T_{uu})du,
\label{ConservedChargef(R)3}
\end{eqnarray}%
where the integrand is precisely $A(u,v)$ in (\ref{3eq6}).
Thus, we have finished our proof of the equivalence of the two
methods.

In addition, the useful components of $K_{\mu \nu }\equiv \nabla
_{\mu }K_{\upsilon }$
and $F^{\mu \nu }\equiv \nabla ^{\mu }\nabla ^{\nu }F$ in coordinates $%
(t,\rho ,\theta ,\phi )$ are
\begin{eqnarray}
K_{tt} &=&e^{-\psi }(\psi ,_{t}r,_{\rho }-r,_{t\rho }),~~K_{t\rho
}=-e^{\psi }r,_{tt},~~K_{\rho t}=-\partial _{\rho }(e^{-\psi
}r,_{\rho })+\psi
,_{t}e^{\psi }r,_{t},\notag \\
K_{\rho \rho }&=&e^{\psi }(\psi _{,t}r_{,\rho }-r_{,t\rho }),~~
F^{tt} =F_{,tt},~~F^{t\rho }=F^{\rho t}=-e^{-2\psi }F_{,\rho
t}+e^{-2\psi
}\psi_{,t}F_{,\rho },\notag \\
F^{\rho \rho }&=&e^{-2\psi }[\partial _{\rho }(e^{-2\psi }F_{,\rho
})+\psi_{,\rho }e^{-2\psi }F_{,\rho }-\psi_{,t}F_{,t}].
\end{eqnarray}
With the help of those quantities,  we can easily check that the constraint
equation (\ref{3eq13}) is satisfied for the FRW universe.

(2) The meaning of the generalized Misner-Sharp energy in FRW
universe.  To see clearly this, let us write down the Friedmann
equations of the $f(R)$ gravity
\begin{eqnarray}
 H^{2}+\frac{k}{a^{2}}
&=&\frac{1}{6f_{R}}[(f_{R}R-f)-6H\partial
_{t}f_{R}+16\pi G \tilde \rho ],  \nonumber  \\
\overset{.}{H}-\frac{k}{a^{2}} &=&\frac{1}{2f_{R}}[H\partial
_{t}f_{R}-\partial _{t}\partial _{t}f_{R}-8\pi G (\tilde \rho
+\tilde p)], \label{5eq4}
\end{eqnarray}
where $\tilde \rho$ and $\tilde p$ are energy density and pressure
of the ideal fluid in the universe. With the first line in
(\ref{5eq4}), we can easily see that the generalized Misner-Sharp
energy in (\ref{4eq8}) can be rewritten as
\begin{equation}
E_{eff}=\tilde \rho V,
 \end{equation}
where $V= 4\pi r^3/3$ is the volume of a sphere with radius $r$.
Therefore, in fact, the generalized Misner-Sharp energy in the FRW
universe is nothing but the total matter energy within a sphere
with radius $r$.

(3) Thermodynamics of apparent horizon in the $f(R)$ gravity. On
the apparent horizon of a FRW universe, the energy crossing the
apparent horizon within time interval $dt$ is~\cite{Cai,Cai-Kim}
\begin{equation}
\delta Q=dE_{eff}|_{r_{A}}=A(\tilde \rho +\tilde p)Hr_{A}dt.
\end{equation}
Note that the horizon entropy of the $f(R)$ gravity is
$S=\frac{A}{4G}f_{R}=\pi r_{A}^{2}f_{R}/G$, while the temperature
of the apparent horizon is~\cite{Cai-Kim,Cai2,Li}:
$T=\frac{1}{2\pi r_{A}}$. Obviously, the usual Clausius relation
$\delta Q= TdS$ does not hold. On the other hand, an internal
entropy production is needed to balance the energy conservation,
$\delta Q=TdS+Td_{i}S$ with
\begin{equation}
d_{i}S=\pi r_{A}[Hr_{A}^{3}(H\partial _{t}f_{R}-\partial
_{t}\partial _{t}f_{R})-\partial _{t}f_{R}r_{A}]/G.
\end{equation}%
This is an effect of the non-equilibrium thermodynamics of
spacetime~\cite{Jacobson1,Cai,Elizalde,Eling}.

(4) The case of scalar-tensor gravity. Indeed the $f(R)$ gravity
is quite similar to scalar-tensor gravity theory in some
sense~\cite{Sotiriou}. Our conclusion on the $f(R)$ gravity
therefore also holds for scalar-tensor gravity. In particular, the
existence of a generalized Misner-Sharp energy has to obey a
constraint condition as well for the scalar-tensor theory.
However, in the FRW universe, a simple expression for the
generalized Misner-Sharp energy can be given, which can be seen in
 appendix~\ref{A}.


\section*{Acknowledgments}
YPH thanks  D. Orlov for useful discussions.  RGC and YPH are
supported partially by grants from NSFC, China (No. 10525060, No.
108215504 and No. 10975168) and a grant from MSTC, China (No.
2010CB833004).
NO was supported in part by the Grant-in-Aid for
Scientific Research Fund of the JSPS No. 20540283,
and also by the Japan-U.K. Research Cooperative Program.
This work is completed during RGC's visit to Kinki
University, Japan with the support of JSPS invitation fellowship.

\appendix

\section{Generalized Misner-Sharp energy of
scalar-tensor theory in FRW universe} \label{A}

The Lagrangian of a generic  scalar-tensor gravity in
4-dimensional space-time can be written as
\begin{equation}
L=\frac{1}{16\pi }F(\phi )R-\frac{1}{2}g^{\mu \nu }\partial_{\mu}\phi
\partial _{\nu }\phi -V(\phi )+L_{m}.  \label{aeq1}
\end{equation}%
where we set Newtonian constant $G=1$, $F(\phi )$ is an arbitrary
positive continuous  function of the scalar field $\phi $, $V(\phi)$
is its potential, and $L_m$ denotes the Lagrangian of other
matter fields. Varying the associated action with respect to
spacetime metric and the scalar field yields equations of motion
\begin{eqnarray}
FG_{\mu \nu }+g_{\mu \nu }\square F-\nabla _{\mu }\nabla _{\nu }F
&=&8\pi
(T_{\mu \nu }^{\phi }+T_{\mu \nu }^{m}),  \label{aeq2} \\
\square \phi -V'(\phi )+\frac{1}{16\pi }F'(\phi )R &=&0.
\label{aeq3}
\end{eqnarray}%
where $T_{\mu \nu }^{m}$ is the energy-momentum tensor of matter fields, and $%
T_{\mu \nu }^{\phi }$ is defined as
\begin{equation}
T_{\mu \nu }^{\phi }=\partial _{\mu }\phi \partial _{\nu }\phi -g_{\mu \nu }\Big(%
\frac{1}{2}g^{\rho \sigma }\partial _{\rho }\phi \partial _{\sigma
}\phi +V(\phi )\Big).
\end{equation}%
Note that here $T_{\mu \nu }^{\phi }$ is not the energy-momentum
tensor of the scalar field.
Similar to the case of f(R) gravity, we  find that the current $%
J^{\mu }=-T_{\nu}^{\mu (m)}K^{\nu}$ is  not always divergence-free
 unless the condition is satisfied
\begin{equation}
(\nabla _{\mu }\nabla _{\nu }F+8\pi \partial _{\mu }\phi \partial
_{\nu }\phi )\nabla ^{\mu }K^{\nu}=0.  \label{aeq5}
\end{equation}%
However, we can easily check that the condition (\ref{aeq5}) can
be always satisfied for the FRW universe (\ref{FRWMetric}) by
using~(\ref{Components}). Some useful components of equations
of gravitational field (\ref{aeq2}) are given by
\begin{eqnarray}
8\pi T_{tt}^{m} &=&3F\Big(\frac{k}{a^{2}}+H^{2}\Big)+3H\overset{.}{F}-8\pi
\Big(\frac{1}{2}\overset{.}{\phi }^{2}+V\Big),~~8\pi T_{t\rho }^{m}=0,  \notag \\
8\pi T_{\rho \rho }^{m} &=&\frac{a^{2}}{1-k\rho ^{2}}\Big[-F\Big(\frac{k}{a^{2}}%
+H^{2}+\frac{2\overset{..}{a}}{a}\Big)-\overset{..}{F}-2H\overset{.}{F}+8\pi \Big(-%
\frac{1}{2}\overset{.}{\phi }^{2}+V\Big)\Big].  \label{aeq6}
\end{eqnarray}%
In this case,  corresponding $A$ and $B$ in (\ref{4eq3}) for the
 scalar-tensor theory, respectively, are
\begin{eqnarray}
A(t,\rho ) &=&\frac{1}{2}Hr^{3}\Big[F\Big(\frac{k}{a^{2}}+H^{2}+\frac{2\overset{..}{a%
}}{a}\Big)+\overset{..}{F}+2H\overset{.}{F}-8\pi \Big(-\frac{1}{2}\overset{.}{\phi }%
^{2}+V\Big)\Big],  \notag \\
B(t,\rho ) &=&\frac{1}{2}\rho ^{2}a^{3}\Big[3F\Big(\frac{k}{a^{2}}+H^{2}\Big)
+3H\overset{.}{F}-8\pi \Big(\frac{1}{2}\overset{.}{\phi }^{2}+V\Big)\Big].
\label{STFRWQuantitiesAB}
\end{eqnarray}%
They obey  the integrable condition (\ref{4eq6}). With these
quantities we can obtain the generalized Misner-Sharp energy of
the scalar-tensor theory in the FRW universe
\begin{equation}
E_{eff}\ =\int B(t,\rho )d\rho =\frac{r^{3}}{2}[F(\frac{k}{a^{2}}+H^{2})+H%
\overset{.}{F}-\frac{8\pi }{3}(\frac{1}{2}\overset{.}{\phi
}^{2}+V)]. \label{EnergyFRWST}
\end{equation}
Comparing this with the first equation in (\ref{aeq6}), one can
immediately see that the generalized Misner-Sharp energy is
just the total matter energy in a sphere with radius $r$
in the FRW universe.


\end{document}